# Reusing Operational Evidence After Context Changes: A Conservative Bayesian Framework for Autonomous Vehicle Safety

Robab Aghazadeh-Chakherlou[1], Siddartha Khastgir[1], Xingyu Zhao[1]
*WMG, University of Warwick, Coventry, UK*
Email: {R.Aghazadeh-Chakherlou, S.Khastgir.1, Xingyu.Zhao}@warwick.ac.uk

**ABSTRACT**

Operational evidence, i.e., evidence of operation without failure is an important component of confidence in the safety or reliability of a system in service, but it is costly to collect. When the context of operation changes, the relevance of previously collected operational evidence becomes unclear. This problem arises, for example, when an autonomous vehicle that has operated safely in one operational environment (the Target Operational Domain, TOD) is deployed in a different but related TOD. Existing practice often treats such evidence in an all-or-nothing manner: either it is fully reused, or it is discarded. Neither position is satisfactory when there are good reasons to believe that the new context is no worse than the previous one, but that belief is itself uncertain.

This paper studies how to make post-change reliability claims by combining evidence from two TODs using Conservative Bayesian Inference (CBI), which combines the evidence from the new context with a weighted amount of previous context and partial prior knowledge through constraints on a set of admissible priors. This yields conservative posterior bounds on quantities of interest. A numerical example illustrates how pre-existing evidence from a previous TOD can be transferred, conservatively and transparently, to support reliability claims in the changed TOD.

## 1. INTRODUCTION

Operational evidence is an important contributor to confidence in the safety or reliability of a system. In autonomous driving, for example, long periods of operation without observed hazardous failure can strengthen assurance arguments about the system's reliability. Such evidence is also explicitly encouraged in standards and guidance, including UL 4600 [1] and IEC 61508 [2], where statistical testing and operational experience support system-level safety claims.

A recurring difficulty, however, is that operational evidence is inherently tied to the context under which it was collected. Here, context includes both the system itself and the conditions under which it operates. Consequently, a change in context may arise from modifications to the system (e.g., software updates, model retraining, or hardware changes) or from changes in the operational environment (e.g., deployment in a different city, changes in traffic patterns, or seasonal variations in weather and illumination). Once such changes occur, the relevance of past evidence becomes uncertain. This issue is particularly important for autonomous vehicles (AV) deployed under changing target operational domains (TODs). A TOD can be understood as a specification of the operational conditions under which the system is intended to function, including environmental, temporal, and traffic-related factors that define the safety-relevant input space. In this paper, we investigate how pre-change operational evidence can be used to support post-change safety claims in a conservative manner.

We focus on changes in context of an AV arising from variations in the TOD, while keeping the system itself fixed. In particular, we consider the case where an autonomous vehicle operates in the same city under different seasonal conditions (changing TOD caused by seasonal variation). Even when the road network and system remain unchanged, the distribution of encountered conditions may vary substantially over time. For example, the frequency of rain, wet roads, darkness, glare, traffic density, and their combinations may differ between autumn–winter and spring–summer operation. Since these factors directly influence hazardous exposure, evidence collected in pre-change TOD (autumn-winter) cannot automatically be taken as equally representative of post-change TOD (spring-summer).

In practice, an assessor may nevertheless have good reasons to believe that a post-change TOD is not worse from a safety point of view than a pre-change one, in the sense that the probability of failure per scenario[1] is no greater. External evidence—such as weather statistics, route characteristics, and traffic data—may indicate that the new TOD contains fewer adverse combinations of safety-relevant conditions, leading to reduced exposure to hazardous scenarios.

This type of reasoning aligns with a broader principle widely used in industrial and regulatory practice: that a modified or newly deployed system should be "No Worse Than the Existing System" (NWTES), as used throughout this paper. Variants of this principle appear across domains, including the "globally at least equivalent" (GALE/GAME) concept in transport safety, "substantial equivalence" in U.S. FDA regulation, and "proven-in-use" arguments in standards such as IEC 61508 [2] and ISO 26262 [3]. While these frameworks aim to ensure that changes do not degrade safety, they do not specify how strongly such prior judgments should influence post-change safety claims when combined with new evidence. As a result, current practice often adopts an implicit "all-or-nothing" approach: either pre-change operational evidence is fully reused, or it is discarded altogether. Both extremes are unsatisfactory. Blind reuse may lead to overconfident claims, while complete rejection wastes valuable information and increases the burden of costly operational testing.

This paper addresses this gap using Conservative Bayesian Inference (CBI) [4, 5], which combines evidence from the pre-change context with weighted post-change evidence, reducing the amount of new evidence required. Instead of specifying a single prior distribution, CBI represents prior knowledge through simple, defensible constraints on a set of admissible priors, avoiding the need to justify a precise prior. In particular, we incorporate the "no-worse-than" (NWTES) assumption [6, 7, 8], which captures the assessor's belief that the post-change context is at least as safe as the pre-change context. This belief is encoded as a constraint on the joint prior, restricting the space of admissible priors. Posterior conclusions are then derived conservatively by optimizing over this set, yielding robust bounds under prior uncertainty.

## 2. METHODOLOGY

Consider an AV operating in a city under two $TOD_A$ (winter) and $TOD_B$ (summer). Let $Y \in \Upsilon$ denote a random variable representing a scenario drawn from the operational space. The space $\Upsilon$ is high-dimensional and may include, among other factors, weather conditions, illumination, traffic characteristics, road geometry, and interactions between road users. Thus, changing the season causes changes in TOD. The $TOD_A$ and $TOD_B$ are represented by two probability distributions over the scenario space. This reflects the fact that a TOD is defined not only by the set of possible scenarios, but also by the frequency with which these scenarios occur. Since the probability of failure depends on exposure to different scenarios, it is necessary to model TODs as distributions over the scenario space rather than as sets: $Y \sim P_A$ in $TOD_A$, $Y \sim P_B$ in $TOD_B$ with $P_A \neq P_B$ in general. As a concrete example, $TOD_A$ may correspond to autumn–winter operation and $TOD_B$ to spring–summer operation in the same city. The road network remains unchanged, but the distributions differ in the frequency of safety-relevant scenarios. In particular, $TOD_B$ may contain fewer adverse scenarios, such as those involving rain, low visibility, or high traffic density under challenging illumination conditions. Let $X_A$ and $X_B$ denote the (unknown) probabilities of failure per scenario in $TOD_A$ and $TOD_B$, respectively, defined as:

$$X_A = \Pr(failure \mid P_A)\ , X_B = \Pr(failure \mid P_B),$$

and treated as random variables to represent epistemic uncertainty about the reliability in each TOD.

[1] A scenario is defined as the temporal evolution of a sequence of scenes, where each scene represents a snapshot of the environment, including all relevant dynamic elements and their relationships. The probability of failure per scenario therefore captures the likelihood that the system fails when exposed to a randomly drawn scenario from the operational distribution.

**Prior knowledge from TOD analysis.** Information about the relationship between $TOD_A$ and $TOD_B$ can be obtained from external evidence, such as: statistical data on weather and illumination conditions, traffic-density measurements, analysis of route usage and exposure, expert assessment of scenario difficulty, etc. This information does not directly specify the values of $X_A$ and $X_B$, but it provides partial knowledge that can be expressed as constraints on their joint prior distribution.

**Marginal constraints.** Another partial knowledge that can be extracted is assuming an upper bound on the probability of failure in each context. The quantities ε and θ encode prior beliefs about acceptable levels of failure probability. $\Pr(X_A \le \epsilon) = \theta,\ \Pr(X_B \le \epsilon) = \theta$ expressing that, with confidence $\theta$, the failure probability in each TOD is below a small reference level $\epsilon$. Such bounds may be obtained from prior assurance arguments, historical operational evidence, or conservative engineering judgments.

**NWTES constraint.** The key TOD-based assumption is that $TOD_B$ is not worse than $TOD_A$ from a safety perspective (i.e., $TOD_B$ is NWTES). This is expressed as: $\Pr(X_B \le X_A) = \phi$ where $\phi$ represents the assessor's confidence in the claim that operation in $TOD_B$ is at least as safe as in $TOD_A$. This confidence can be informed quantitatively by comparing the scenario distributions $P_A$ and $P_B$. In particular, if $Y_{adv} \subset \Upsilon$ denotes the set of adverse or safety-critical scenarios, then $P_B(Y_{adv}) < P_A(Y_{adv})$ indicates reduced exposure in $TOD_B$, supporting the NWTES claim and providing a basis for assigning ϕ.

**Operational evidence.** Operational evidence is obtained by observing the system over $n_A$ and $n_B$ scenarios in $TOD_A$ and $TOD_B$, respectively, with no observed failures. This evidence is combined with the prior constraints using Bayesian inference.

***Theorem 1:*** Let $\mathcal{D}$ be the set of all joint prior distributions $F_{AB}$ of $(X_A, X_B)$ defined on $[0,1] \times [0,1]$, satisfying the TOD-based constraints described above. Assume that $n_A$ and $n_B$ failure-free scenarios have been observed in $TOD_A$ and $TOD_B$, respectively. Then the conservative posterior confidence $C$ in the claim $X_B \le p_r$ ($p_r$ is a threshold) is obtained as the solution of the following optimization problem[2]:

$$\begin{cases} C = \inf_{\mathcal{D}} \Pr(X_B \le p_r \mid n_A, n_B) = \inf_{\mathcal{D}} \dfrac{\int_{[0,1]\times[0,p_r]} (1-x_A)^{n_A}(1-x_B)^{n_B}\, dF_{AB}(x_A, x_B)}{\int_{[0,1]\times[0,1]} (1-x_A)^{n_A}(1-x_B)^{n_B}\, dF_{AB}(x_A, x_B)}, \\ \text{s.t.} \\ \Pr(X_A \le \epsilon) = \theta,\ \ \epsilon < p_r \\ \Pr(X_B \le \epsilon) = \theta,\ \ \epsilon < p_r \\ \Pr(X_B \le X_A) = \phi \end{cases} \tag{1}$$

## 3. Numerical Example

The theorem is applied with: ε = $10^{-4}$, p_r = $10^{-3}$, θ = 0.8, and ϕ = 0.9, where ϕ is motivated by the reduced adverse scenario exposure in $TOD_B$: $P_A(Y_{adv}) = 0.14$ versus $P_B(Y_{adv}) = 0.09$. Operational evidence ranges are $n_A = 0 \ldots 1000$ and $n_B = 0 \ldots 500$. Figure (1a) shows that even at $n_B = 0$, the posterior confidence already exceeds the prior θ = 0.8, because $n_A = 500$ failure-free scenarios in $TOD_A$ contribute indirectly through the NWTES constraint. Confidence then increases monotonically with

[2] The proof of the theorem is omitted here for brevity. A similar proof can be found in [6 - 8].

$n_B$ , with diminishing returns. Figure (1b) shows the effect of $n_A$ with $n_B = 200$ fixed: confidence rises from 0.844 at $n_A = 0$ to approximately 0.899 at $n_A = 1000$, saturating as the worst-case prior concentrates on a fixed structure. The limiting value is determined by θ, ϕ, ε, and $p_r$ , not by $n_A$ alone. Figure (1c) shows the combined effect: for a fixed $n_B$ , larger $n_A$ raises the baseline confidence uniformly, while for a fixed $n_A$ , increasing $n_B$ dominates the growth. Overall, $TOD_A$ evidence can partially substitute for $TOD_B$ evidence — 1000 failure-free scenarios in $TOD_A$ substitute for approximately 120 direct scenarios in $TOD_B$ — reducing the new testing required in the target domain.

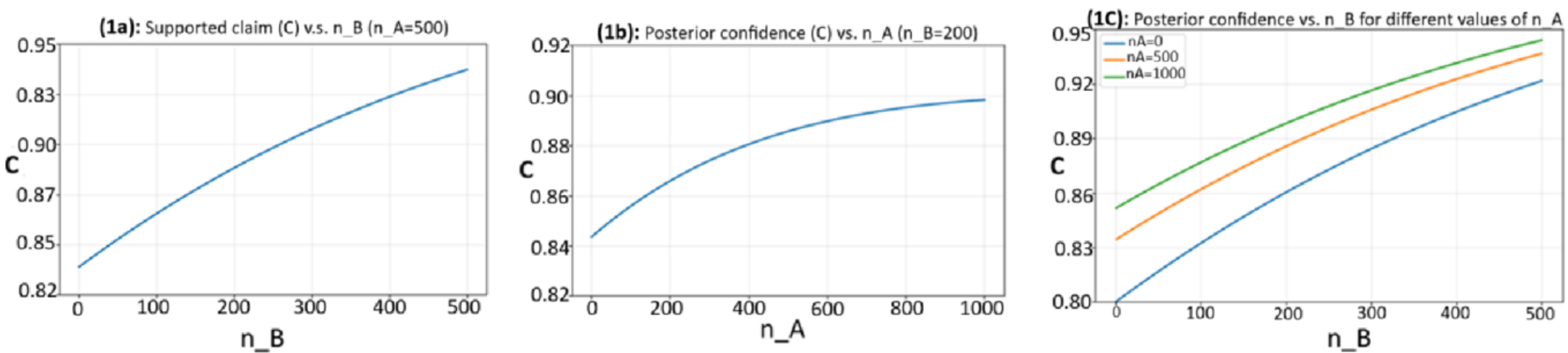

*Figure 1: Conservative posterior confidence (C): 1a) C vs.* $n_B = 0 \dots 500$ *for a fixed* $n_A = 500$ *, 1b) C vs.* $n_A = 0 \dots 1000$ *for a fixed* $n_B = 200$ *, 1c) C vs.* $n_B = 0 \dots 500$ *for different values of* $n_A = 0, 500, 1000$

## 4. Discussion and Conclusion

This paper presented a CBI-based framework for making conservative post-change reliability claims when an autonomous vehicle is redeployed in a new TOD. By encoding the no-worse-than (NWTES) belief as a prior constraint, the framework enables principled transfer of operational evidence across contexts, reducing the need for extensive post-change testing. The results show that pre-change evidence provides a meaningful but saturating improvement in posterior confidence, while maintaining conservatism through worst-case prior assumptions. The posterior confidence is guaranteed to be a lower bound over all priors consistent with the stated constraints, making it suitable for regulatory arguments under standards such as IEC 61508 [2] and ISO 26262 [3].

Future work will extend the framework to account for changes in the system itself, such as software updates, alongside TOD changes. Integrating the scenario-distribution representativeness metric of [8] into the CBI formulation would provide a unified pipeline in which TOD comparison grounds the NWTES confidence parameter ϕ, and uncertainty in operational exposure propagates directly into the reliability claim.